\documentclass[aps,pra,twocolumn,superscriptaddress,showpacs,floatfix,longbibliography]{revtex4-1}
\usepackage{bm}
\usepackage{amssymb}
\usepackage{epsfig}
\usepackage{amsmath}
\usepackage{color}
\usepackage{graphicx}
\usepackage{epstopdf}
\usepackage[toc]{appendix}

\begin{document}

\title{Superscattering of pseudospin-1 wave in photonic lattice}

\author{Hongya Xu}
\affiliation{School of Electrical, Computer, and Energy Engineering,
Arizona State University, Tempe, AZ 85287, USA}

\author{Ying-Cheng Lai} \email{Ying-Cheng.Lai@asu.edu}
\affiliation{School of Electrical, Computer, and Energy Engineering,
Arizona State University, Tempe, AZ 85287, USA}
\affiliation{Department of Physics, Arizona State University, Tempe,
AZ 85287, USA}

\date{\today}

\begin{abstract}

We uncover a superscattering behavior of pseudospin-1 wave from weak
scatterers in the subwavelength regime where the scatterer size is much
smaller than wavelength. The phenomenon manifests itself as unusually
strong scattering characterized by extraordinarily large values of the
cross section even for arbitrarily weak scatterer strength. We establish
analytically and numerically that the physical origin of superscattering
is revival resonances, for which the conventional Born theory breaks down.
The phenomenon can be experimentally tested using synthetic photonic
systems.

\end{abstract}

\maketitle

\section{Introduction} \label{sec:intro}

In wave scattering, a conventional and well accepted notion is that
weak scatterers lead to weak scattering. This can be understood by
resorting to the Born approximation. Consider a simple
2D setting where particles are scattered from a circular
potential of height $V_0$ and radius $R$. In the low energy (long wavelength)
regime $kR < 1$ (with $k$ being the wavevector), the Born approximation holds
for weak potential: $(m/\hbar^2)|V_0|R^2 \ll 1$. Likewise, in the
high energy (short wavelength) regime characterized by $kR > 1$, the Born
approximation still holds in the weak scattering regime:
$(m/\hbar^2)|V_0|R^2 \ll (kR)^2$. 
In general, whether scattering is weak or
strong can be quantified by the scattering cross section. For scalar waves
governed by the Schr\"{o}dinger equation, in the Born regime the scattering
cross section can be expressed as polynomial functions of the effective
potential strength and size~\cite{Schiff:book}. For spinor waves described
by the Dirac equation (e.g., graphene systems), the
2D transport cross section is given by~\cite{Wu2014}
$\Sigma_{tr}/R \simeq (\pi^2/4)(V_0R)^2 (kR)$ (under $\hbar v_F=1$). In light
scattering from spherically dielectric, ``optically soft'' scatterers with
relative refractive index $n$ near unity, i.e., $kR|n-1|\ll1$, the Born
approximation manifests itself as an exact analog of the Rayleigh-Gans
approximation~\cite{newton1982}, which predicts that the scattering
cross section behaves as $\Sigma/(\pi R^2)\sim|n-1|^2(kR)^4$ in the small
scatterer size limit $kR\ll 1$. In wave scattering, the conventional
understanding is then that a weak scatterer leads to a small cross section
and, consequently, to weak scattering, and this holds regardless of nature
of the scattering particle/wave, i.e., vector, scalar or spinor.

In this paper, we report a counterintuitive phenomenon that
defies the conventional wisdom that a weak scatterer always results in
weak scattering. The phenomenon occurs in scattering of higher
spinor waves, such as pseudospin-1 particles that can arise in experimental
synthetic photonic systems whose energy band structure consists of a pair of
Dirac cones and a flat band through the conical intersection
point~\cite{Huang2011,Moitra2013,Li2015,Guzman:2014,Mu2015,Vic2015,
Diebel:2016,Taiee2015}. Theoretically, pseudospin-1 waves are effectively
described by the generalized Dirac-Weyl
equation~\cite{Shen:2010,Guzman:2014,Fang2016}:
$H_0\Psi=\boldsymbol{S\cdot k}\Psi=E\Psi$ with
$\Psi=[\Psi_1, \Psi_2, \Psi_3]^T$, $\boldsymbol{k}=(k_x, k_y)$ and
$\boldsymbol{S}=(S_x, S_y)$ being the vector of $3\times 3$ matrices for
spin-$1$ particles. Investigating the general scattering of pseudospin-$1$
wave, we find the surprising and counterintuitive
phenomenon that extraordinarily strong scattering, or superscattering, can
emerge from arbitrarily weak scatterers at sufficiently low energies
(i.e., in the deep subwavelength regime). Accompanying this phenomenon is a
novel type of resonances that can persist at low energies for weak scatterers.
We provide an analytic understanding of the resonance and derive formulas
for the resulting cross section, with excellent agreement with results from
direct numerical simulations. We also propose experimental verification
schemes using photonic systems.

\section{Results} \label{sec:results}

We consider scattering of 2D pseudospin-1 particles from a circularly symmetric
scalar potential barrier of height $V_0$ defined by $V(r) = V_0\Theta(R - r)$,
where $R$ is the scatterer radius and $\Theta$ denotes the Heaviside function.
The band structure of pseudospin-1 particles can be illustrated using a 2D
photonic lattice for transverse electromagnetic wave with the electric
field along the $z$-axis. As demonstrated in previous
works~\cite{Huang2011,Fang2016}, Dirac cones induced by accidental degeneracy
can emerge at the center of the Brillouin zone for proper material
parameters, about which three-component structured light wave emerges
and is governed by the generalized Dirac-Weyl equation.

\begin{figure}[h]
\centering
\fbox{\includegraphics[width=\linewidth]{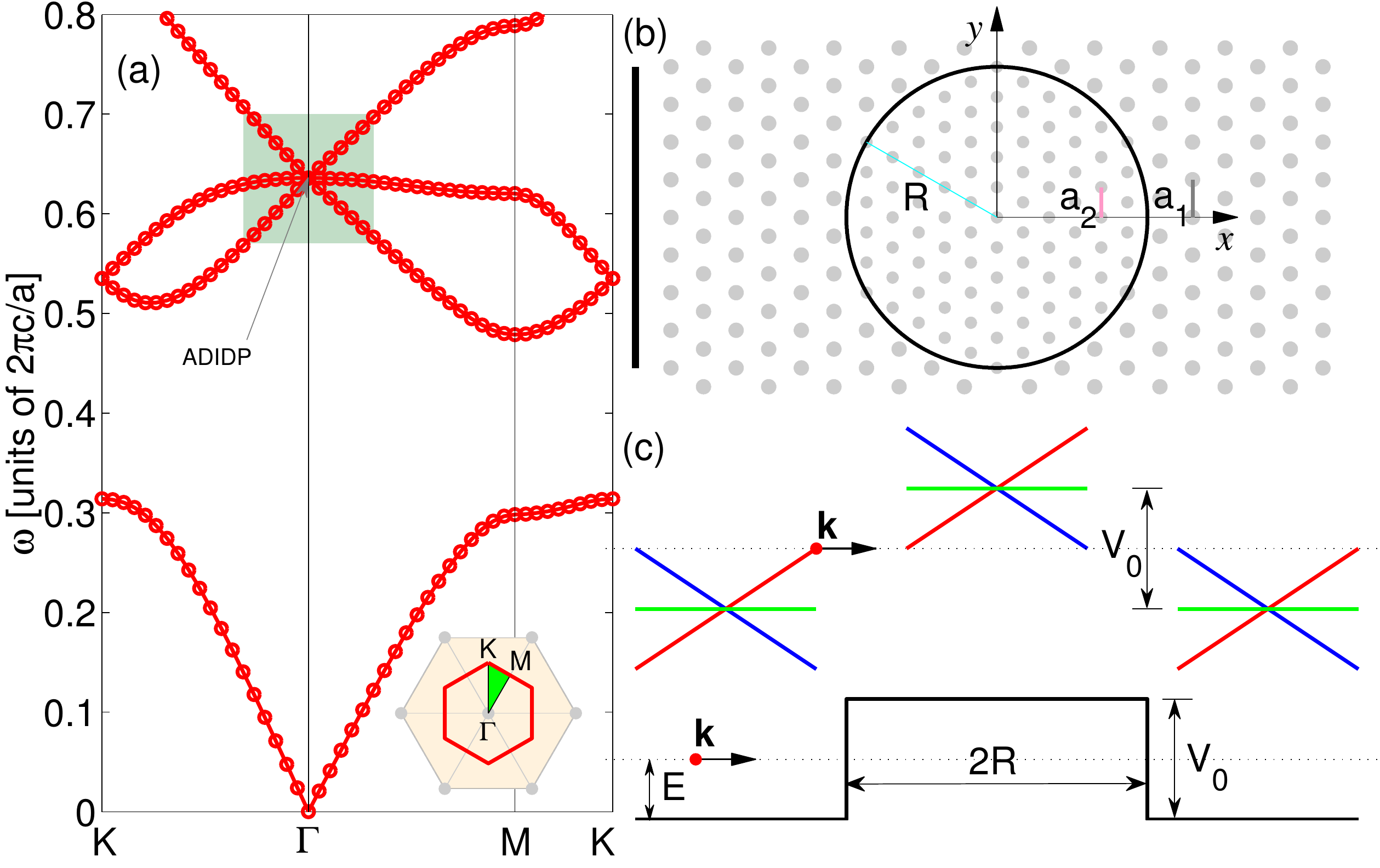}}
\caption{ \textbf{Pseudospin-1 band structure and the underlying photonic
lattice structure}. The lattice has a triangular configuration constructed
by cylindrical alumina rods in air. (a) The band structure with an
accidental-degeneracy induced Dirac point at the center of the 1st
Brillouin zone, (b) sketch of the physical lattice, and (c) band
structures outside and inside of the scatterer. A possible experimental
parameter setting is $a_1 = 17$mm, $r_1 = 0.203a_1$, $a_2 = 0.8a_1$,
and $r_2 = 0.203a_2$. Dielectric constant of alumina rod is 8.8.}
\label{fig:band}
\end{figure}

We consider the setting of photonic crystal to illustrate the
pseudospin-1 band structure. Figure~\ref{fig:band}(a) shows the band
structure of lattices with a triangular configuration constructed by
cylindrical alumina rods in air, where the rod radius 
is $r_0 = 0.203a$ ($a$ - lattice constant) and the rod dielectric constant 
is 8.8~\cite{Huang2011}. We obtain an accidental-degeneracy
induced Dirac point at the center of the 1st Brillouin zone at the finite
frequency of $0.6357\cdot 2\pi \cdot c/r_0$. Following a general lattice
scaling scheme of photonic gate potential~\cite{Fang2016}, we obtain
a sketch of the cross section of the lattice in the plane, as shown in
Fig.~\ref{fig:band}(b), where the thick black bar denotes an applied
exciter. For our scattering problem, the band structures
outside and inside of the scatterer are shown in Fig.~\ref{fig:band}(c).

The scattering problem can be treated analytically using the Dirac-Weyl
equation (see Appendix A for a detailed derivation of the various scattering
formulas). To demonstrate the phenomenon of superscattering, we use the
transport cross section $\Sigma_{tr}$ to characterize the scattering
dynamics. 
(It should be noted that the total cross section $\Sigma$ is 
another usual quantity for characterizing superscattering with consistent 
results as from the transport cross section - see Appendix B for details.) 
The transport cross section is defined in terms of the scattering 
coefficients $A_l$ as:
\begin{equation} \label{eq:sTrCS}
\Sigma_{tr}/R = (4/x) \sum_{l=-\infty}^{\infty}
\left\{|A_l|^2 - \Re\left[A_l(A_{l+1})^*\right]\right\},
\end{equation}
where $A_l$'s can be obtained through the standard method of partial wave
decomposition~\cite{Schiff:book}. For convenience, we define
$\rho \equiv V_0R$ and $x \equiv kR$.
At low energies, i.e., $x \ll 1$, scattering is dominated by the lowest
angular momentum channels $l=0, \pm1$. To reveal the relativistic quantum
nature of the scattering process, we focus on the under-barrier scattering
regime, i.e., $x<\rho$, so that manifestations of phenomena such as Klein
tunneling are pronounced. We define two subregimes of low energy
scattering: $1<\rho$ and $x<\rho<1$, where the former corresponds
to the case of a scatterer with a large scattering potential. 
The weak scatterer subregime, i.e., $x<\rho<1$, is one in which the 
counterintuitive phenomenon of superscattering arises. Specifically, 
for $x<\rho<1$, we obtain the leading coefficients as 
\begin{eqnarray}
A_0 & \approx & -P_0/(P_0 + iQ_0) \\ \nonumber
A_{\pm1} & \approx & -P_1/[P_1 + i(4 + Q_1)], 
\end{eqnarray}
where $P_0 = \pi x$ and
\begin{displaymath}
Q_0 = 2\left(x\ln{(\gamma_Ex/2)} - J_0(\rho-x)/[J_1(\rho-x)]\right)
\end{displaymath}
with $\ln\gamma_E\approx0.577\cdots$ being the Euler's constant and
$P_1,Q_1$ given by relations $[P_1 , Q_1]= x[P_0, Q_0]$. Using these
relations, we obtain
\begin{equation} \label{eq:Tr}
\frac{\Sigma_{tr}}{R} = \frac{4P_0^2}{x(P_0^2 + Q_0^2)}
\left\{1 - \frac{8Q_1}{P_1^2 + (4 + Q_1)^2}\right\}.
\end{equation}
We first show that, in our scattering system, all the conventional
resonances will disappear in the weak scatterer regime ($\rho<1$). To make
an argument, we examine the case of a scatterer with 
large scattering potential: $\rho > 1$ where
the transport cross section as a function of $x$ and $\rho$ is given by 
(see Appendix A for a detailed derivation)
\begin{eqnarray} \label{eq:TrR}
\frac{\Sigma_{tr}}{R} & \approx & \frac{4}{x}
\left(\frac{(\pi x)^2}{(\pi x)^2+4[\rho-\rho_{0,m}+x\ln(\gamma_Ex/2)]^2}\right) 
\\ \nonumber
& + & \frac{8}{x}\left( \frac{(\pi x^3)^2}{(\pi x^3)^2 
+ 4(\rho-\rho_{1,n}-x)^2}\right),
\end{eqnarray}
with $m,n=1,2,3,\cdots$ and $\rho_{0,m}, \rho_{1, n}$ denoting the $m$th and 
$n$th zeros of the Bessel functions $J_{0}$ and $J_1$, respectively.
The resonances occur about
$\rho \approx \rho_{0,m}, \rho_{1,n}$ for $x\ll 1$,
and thus are well separated with a minimum position at $\rho \approx 2.4$.
This indicates that the locations of such resonances 
satisfy $\rho > 2$, which are not possible in the small scattering 
potential regime $\rho < 1$. In conventional scattering
systems where the Born approximation applies, no additional resonances
will emerge in the small scattering potential regime $\rho<1$.

\begin{figure}[h]
\centering
\fbox{\includegraphics[width=\linewidth]{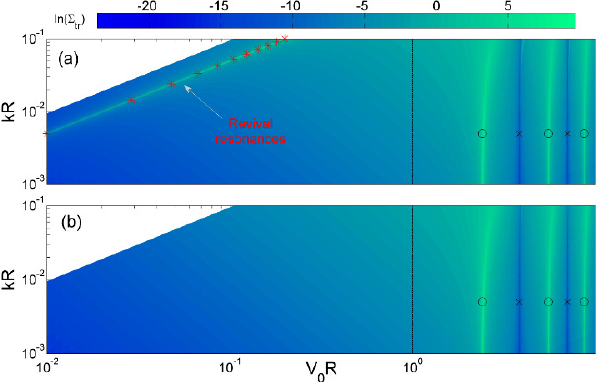}}
\caption{ \textbf{Persistent revival resonances of pseudospin-1 particles 
from a weak circular scatterer at low energies}. (a) Contour map of transport
cross section \textcolor{blue}{in unit of $1/R$} (on a logarithmic scale) versus the scatterer strength
$\rho=V_0R$ and size $x=kR$ for relativistic quantum scattering of
2D massless pseudospin-$1$ particles. Revival resonances occur, which can
lead to superscattering (see Fig.~\ref{fig:SSFig2} below).
(b) Similar plot for pseudospin-$1/2$ particles for comparison,
where no resonances occur, implying total absence of superscattering.
The scatterer is modeled as a circular step like potential
$V(r)=V_0\Theta(R-r)$, representing a finite size scalar impurity or
an engineered scalar-type of scatterers. The markers correspond to
the theoretical prediction, where the black circles ($\circ$)
and crosses ($\times$) are from $\rho\approx\rho_{0,m},\rho_{1,n}$ (for $x\ll 1$),
and the red stars (\textcolor[rgb]{1,0,0}{$*$}) follow the revival resonant
condition given by $\rho = 2x$ for $\rho\ll 1$.}
\label{fig:SSFig1}
\end{figure}

For sufficiently weak scatterer strength ($\rho\ll 1$), the prefactor in
\eqref{eq:Tr}, i.e.,
\begin{eqnarray}
\nonumber
4P_0^2/[x(P_0^2 + Q_0^2)] & \approx & \pi^2J_1^2(\rho-x)x/[J_0^2(\rho-x)]
\\ \nonumber
&\rightarrow & (\pi^2/4)(\rho-x)^2 x\ll1,
\end{eqnarray}
is off-resonance. The remaining factor characterizes the emergence of
a new type of (unconventional) revival resonances at $Q_1+4=0$,
which are unexpected as the scatterers are
sufficiently weak so, according to the conventional Born theory, no
scattering resonances are possible. The resonant condition can be
obtained explicitly from the constraint
\begin{displaymath}
Q_1 + 4 =0 \Rightarrow xJ_0(\rho-x)=2J_1(\rho-x). 
\end{displaymath}
We obtain
$\rho = 2x$ for $\rho\ll 1$.
The surprising feature of revival resonance is that it persists no matter how
weak the scatterer. As a result, superscattering can occur for arbitrarily
weak scatterer strength. One example is shown in Fig.~\ref{fig:SSFig1}(a),
where a good agreement between the theoretical prediction and numerical
simulation is obtained. For comparison, results for the corresponding
pseudospin-$1/2$ wave scattering system governed by the conventional
massless Dirac equation are shown in Fig.~\ref{fig:SSFig1}(b), where
scattering essentially diminishes for near zero scatterer strength,
indicating complete absence of superscattering.

\begin{figure}[h]
\centering
\fbox{\includegraphics[width=\linewidth]{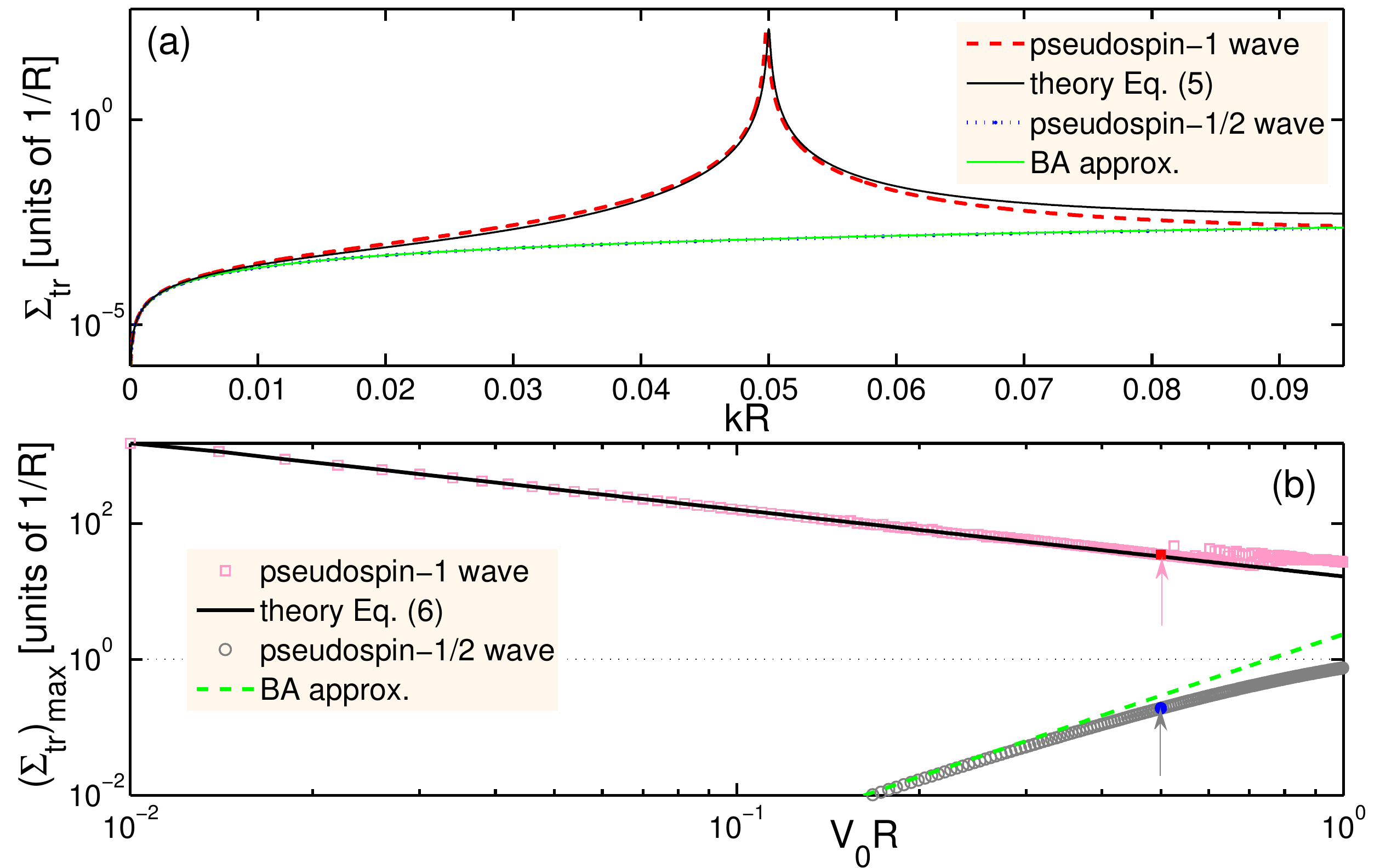}}
\caption{ \textbf{Superscattering of pseudospin-1 wave.}
(a) Transport cross section as a function of $x\equiv kR$ for a weak
scatterer of strength $\rho\equiv V_0R=0.1$, and (b) dependence of the
maximum transport cross section on $V_0R$.}
\label{fig:SSFig2}
\end{figure}

To characterize superscattering in a more quantitative manner, we obtain
from Eq.~(\ref{eq:Tr}) the associated resonance width as
$\Gamma\sim\pi\rho^3/8$, and the closed approximation form as 
\begin{equation}
\frac{\Sigma_{tr}}{R}\approx \frac{\pi^2}{4}\rho^2x\left[1 + \frac{16x\rho}{\pi^2x^4\rho^2 + 16(\rho-2x)^2}\right].
\end{equation}
In addition, at the resonance, we have 
\begin{equation} \label{eq:maxRCS}
\left(\frac{\Sigma_{tr}}{R}\right)_{\textrm{max}}\approx\left.
\frac{\pi^2xJ_1^2(x)}{J_0^2(x)}\frac{32}{\pi^2x^4}\right|_{x=\rho/2}
\simeq\frac{16}{\rho}.
\end{equation} 
A striking and counterintuitive consequence of \eqref{eq:maxRCS} is
that, the weaker the scatterer ($\rho\downarrow$), the larger the
resulting maximum cross section ($(\Sigma_{tr}/R)_{\textrm{max}}\uparrow$).
This can be explained by noting that, due to the revival resonant scattering,
an arbitrarily large cross section can be achieved for a sufficient weak
scatterer with its radius $R$ much smaller than the incident wavelength
$2\pi/k$ (i.e., in deep-subwavelength regime $kR\ll 1$). In contrast,
for a system hosting pseudospin-$1/2$ wave under the same condition of
$x<\rho\ll 1$ where the Born approximation applies~\cite{Wu2014},
the maximum transport cross section is given by
\begin{equation} \label{eq:maxCS}
\left(\frac{\Sigma_{tr}}{R}\right)_{\textrm{max}}^{BA}
\approx\frac{\pi^2}{4}\rho^3.
\end{equation}
Comparing with pseudospin-$1/2$ particles, the scattering behavior
revealed by Eq.~(\ref{eq:maxRCS}) for pseudospin-1 particles is extraordinary
and represents a fundamentally new phenomenon which, to our knowledge,
has not been reported for any wave (especially matter wave) systems. The
analytic predictions [Eqs.~(\ref{eq:maxRCS}) and (\ref{eq:maxCS})] have
been validated numerically, as shown in Fig.~\ref{fig:SSFig2}.

\begin{figure}[h]
\centering
\fbox{\includegraphics[width=\linewidth]{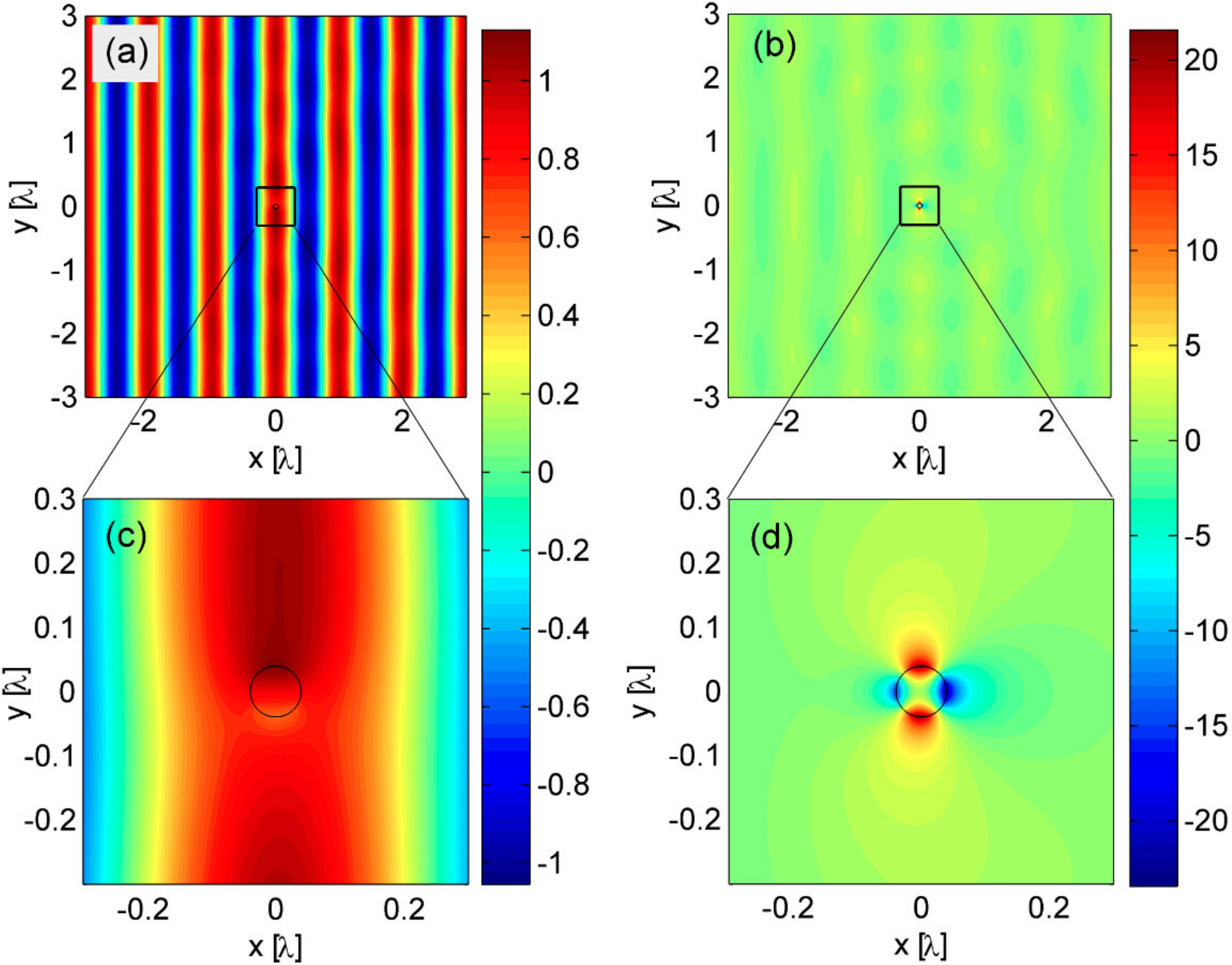}}
\caption{ \textbf{Wavefunction patterns associated with superscattering.}
For $V_0R=0.5$ and $kR=0.2485$, distribution of the real part of one
component of the spinor wavefunction $\Re{(\Psi_2)}$ for
(a) pseudospin-$1/2$ and (b) pseudospin-$1$ wave. (c,d) Magnification
of part of (a) and (b), respectively. \textcolor{blue}{Both axies in (a)-(d) are in unit of the incident wavelength $\lambda$. The color code denotes the quantity $\Re{(\Psi_2)}$.}}
\label{fig:SSFig3}
\end{figure}

Further insights into superscattering can be obtained by examining the
underlying wavefunction patterns, as shown in Fig.~\ref{fig:SSFig3}.
In particular, Figs.~\ref{fig:SSFig3}(a,c) and \ref{fig:SSFig3}(b,d)
show the distributions of the real part of one component of the spinor
wavefunction $\Re{(\Psi_2)}$ for pseudospin-$1/2$ and pseudospin-1
particles, respectively, where the parameters are $V_0R=0.5$ and $kR=0.2485$.
The patterns in Figs.~\ref{fig:SSFig3}(b,d) correspond to the revival
resonance indicated by the pink arrow in Fig.~\ref{fig:SSFig2}(b).
We see that, even for such a weak scatterer, the incident pseudospin-1
wave of a much larger wavelength $\lambda=2\pi/k\sim 25R$ is effectively
blocked via trapping around the scatterer boundary, resulting in strong
scattering. In contrast, for the conventional pseudospin-$1/2$ wave
system, the weak scatterer results in only weak scattering, as shown in
Figs.~\ref{fig:SSFig3}(a,c), which is anticipated from the Born theory.

\section{Experimental test with photonic systems} \label{sec:exp_proposal}

It is possible to test superscattering in experimental optical systems.
Recent realization of photonic Lieb lattices consisting of evanescently 
coupled optical waveguides implemented by femtosecond laser-writing 
technique~\cite{Guzman:2014,Mu2015,Vic2015,Diebel:2016} make them suitable 
for studying the physics of pseudospin-$1$ Dirac cones. 
For example, in the tight-binding framework, for a homogeneous identical 
waveguide array with the same propagation constant $\beta_0$, the 
Hamiltonian in the momentum space is given by
\begin{equation}
\mathcal{H}_{TB}(\boldsymbol{k}) = \begin{pmatrix}
\beta_0 & 2\kappa_x\cos(\frac{k_x}{2}) & 0 \\
2\kappa_x\cos(\frac{k_x}{2}) & \beta_0 & 2\kappa_y\cos(\frac{k_y}{2}) \\
0 & 2\kappa_y\cos(\frac{k_y}{2}) & \beta_0
\end{pmatrix}. 
\end{equation}
In the low-energy regime (measured from the $\beta_0$),
the Hamiltonian is reduced to a generalized Dirac-Weyl Hamiltonian for
spin-$1$ particles with $\beta_0$ analogous to the constant electronic
gate (voltage) potential. As such, the superscattering phenomenon uncovered
in our work can in principle be tested experimentally in photonic Lieb
lattice systems through a particular design of the refractive index 
profile across the lattice to realize the scattering configuration.

Loading ultracold atoms into an optical
Lieb lattice fabricated by interfering counter-propagating laser
beams~\cite{Taiee2015} provides another versatile platform to test our
findings, where appropriate holographic masks can be used to implement
the desired scattering potential barrier~\cite{Bakr2009,Dora2011}.
Synthetic photonic crystal based 2D pseudospin-$1$ wave systems are also
promising for feasible experimental validation. For example, it was
demonstrated experimentally~\cite{Huang2011,Moitra2013,Li2015} and
theoretically~\cite{Mei:2012,Fang2016} that a pseudospin-$1$ wave system
can be realized in 2D dielectric photonic crystals
via the principle of accidental degeneracy. Implementation of the scalar
type of potential can be achieved by manipulating the length scale of the
photonic crystals. From a recent work of ``on-chip zero-index metamaterial''
design~\cite{Li2015} based on such a system, we note that the phenomenon
of superscattering uncovered in this paper can be relevant to a novel
on-chip superscatterer fabrication, which is not possible for conventional
wave systems.

\section{Conclusion and discussion} \label{sec:conclusion}

In conclusion, we uncover a superscattering phenomenon in a class of $2$D wave
systems that host massless pseudospin-$1$ particles described by the
Dirac-Weyl equation, where extraordinarily strong scattering (characterized
by an unusually large cross section) occurs for arbitrarily
weak scatterer in the low energy regime. Physically, superscattering can be
attributed to the emergence of persistent revival resonances for
scatterers of weak strength, to which the cross section is inversely
proportional. These unusual features defy the prediction of the Born
theory that is applicable but to conventional electronic or optical
scattering systems. Superscattering of pseudospin-$1$ wave thus represents
a fundamentally new scattering scenario, and it is possible to conduct
experimental test using synthetic photonic systems.

An important issue is whether superscattering uncovered in this paper is
due to the presence of a flat band that implies an infinite density of
states. Our answer is negative, for the following reasons. Note that,
measured from the three-band intersection point, the energy for the
(dispersionless) flat band states is zero outside and $V_0$ inside the
scatterer, but for the two dispersion Dirac bands the energy is finite
outside the scatterer and not equal to $V_0$ inside. For elastic
scattering considered in our work, the incident energy outside
the scatterer is finite and less than $V_0$ as well. As a result, only
the states belonging to the conical dispersion bands are available both
inside and outside the scatterer, and therefore are responsible for the
superscattering phenomenon. Indeed, as demonstrated, superscattering is
due to revival resonant scattering for states belonging to the conical
dispersion bands that persist in the regime of arbitrarily weak scatterer
strength. From another angle, if superscattering were due to the flat band,
the phenomenon would arise in the conventional resonant scattering regime
$V_0R > 1$, which has never been observed.

While the flat band itself is not directly relevant to the
superscattering behavior, its presence makes the structure of the relevant
states belonging to the conical bands different from those, e.g., in a two
band Dirac cone system, giving rise to boundary conditions that permit
discontinuities in the corresponding intensity distribution and tangent
current at the interface. Interestingly, surface plasmon modes
[c.f., Fig.~\ref{fig:SSFig3}(d)] are excited at the interface when
revival resonant scattering occurs, which are strongly localized and
can be excited for arbitrarily weak scatterer strength, leading to
superscattering in the deep sub-wavelength regime. These modes are
created from the particular spinor structure of the photon states, which
can be implemented by engineering light propagation in periodically
modulated/arranged, conventional dielectric materials (e.g., alumina)
rather than within the material itself. Our finding of the superscattering
phenomenon is thus striking and represents a new scattering capability
that goes beyond the Rayleigh-Gans limit or, equivalently, one defined
by the Born approximation.

With respect to potential applications of the finding of this paper,
it is worth emphasizing that the
phenomenon of superscattering represents a novel way of controlling light
behaviors beyond those associated with the conventional scattering scenario
because, in our system [e.g., Fig.~\ref{fig:band}(b)], light is structured
into three-component spinor states and behaves as relativistic spin-1 wave
in the underlying photonic lattice. There have been extensive recent
experimental works demonstrating that such lattice systems can actually
be realized. Our theoretical prediction is based on a general setting
that effectively characterizes the low-energy physics underlying the
photonic lattices.

\section*{Acknowledgement}
The authors would like to acknowledge support from the Vannevar Bush 
Faculty Fellowship program sponsored by the Basic Research Office of 
the Assistant Secretary of Defense for Research and Engineering and 
funded by the Office of Naval Research through Grant No.~N00014-16-1-2828.
This work was also supported by AFOSR under Grant No.~FA9550-15-1-0151.

\appendix
\appendixpageoff

\section{Scattering formalism of $2$D massless spin-$1$ particles}

\paragraph*{Model Hamiltonian.}
As indicated in the main text, we consider the following perturbed Hamiltonian
\begin{equation}\label{eq:H}
\mathcal{H} = H_0 + V(\boldsymbol{r}),
\end{equation}
where $V(\boldsymbol{r})=V_0\Theta(R-r)$ with $V_0$ being the potential height.

Generally, far away from the scattering center (i.e. $r\gg R$), for an incoming flux along the $x$ direction, the spinor wavefunction with the band index $s$ takes the asymptotic form
\begin{equation}\label{eq:fF1}
|\Psi_{s}^{\gg}(\boldsymbol{r})\rangle = e^{ikx}|\boldsymbol{k}_0,s\rangle + \frac{f(\theta)}{\sqrt{-ir}}e^{ikr}|\boldsymbol{k}_{\theta},s\rangle,
\end{equation}
where the vector $|\boldsymbol{k},s\rangle$ is the spinor plane wave amplitude with wavevectors $\boldsymbol{k}_0=(k, 0)$ and $\boldsymbol{k}_\theta = k(\cos\theta, \sin\theta)$ defining directions of the incident and scattering respectively.

In our case, for the conical dispersion bands $s=\pm$, we obtain
\begin{equation}
|\boldsymbol{k},s\rangle = \frac{1}{2}\begin{pmatrix}
 e^{-i\theta}\\
\sqrt{2}s \\
 e^{i\theta}
\end{pmatrix}.
\end{equation}
With the definition of the current operator $\hat{\boldsymbol{J}}=(1/\hbar)\nabla_{\boldsymbol{k}}H(\boldsymbol{k})=v_F(S_x, S_y)$, we have the scattered current
\begin{equation}
J_{sc} =\frac{1}{r}\langle\boldsymbol{k}_\theta,s|f^*\hat{\boldsymbol{J}}\cdot\frac{\boldsymbol{k}_\theta}{k}f|\boldsymbol{k}_\theta,s\rangle =\frac{v_F}{r}|f(\theta)|^2,
\end{equation}
while the incident current $J_{in}=\langle\boldsymbol{k}_0,s|\hat{\boldsymbol{J}}\cdot\boldsymbol{k}_0/k|\boldsymbol{k}_0,s\rangle=v_F$. The differential cross section is thus defined in terms of the scattering amplitudes $f(\theta)$ as
\begin{equation}
\frac{d\Sigma}{d\theta} = \frac{rJ_{sc}}{J_{in}} = |f(\theta)|^2.
\end{equation}
The other relevant cross sections can be calculated by definition, i.e. the total cross section (TCS)
\begin{equation}\label{eq:TCS}
\Sigma = \int_0^{2\pi}d\theta\frac{d\Sigma}{d\theta},
\end{equation}
the transport cross section (TrCS)
\begin{equation}\label{eq:TrCS}
\Sigma_{tr} = \int_0^{2\pi}d\theta (1-\cos\theta)|f(\theta)|^2.
\end{equation}

In order to figure out the exact expression of $f(\theta)$, we expand the wavefunctions inside and outside the scatterer as a superposition of partial waves, i.e.
\begin{subequations}
for $r>R$ (outside the scatterer)
\begin{equation}\label{eq:OW}
        |\Psi_{s}^{>}(\boldsymbol{r})\rangle = \sum_l\psi_{l,s}^{>}(\boldsymbol{r}),
\end{equation}
for $r<R$ (inside the scatterer)
\begin{equation}\label{eq:IW}
|\Psi_{s}^{<}(\boldsymbol{r})\rangle = \sum_l\psi_{l,s}^{<}(\boldsymbol{r}),
\end{equation}
\end{subequations}
where $\psi_{l,s}^>$ and $\psi_{l,s}^<$ are the partial waves defined in terms of the cylindrical wave eigenfunctions of the reduced Hamiltonian $\mathcal{H}$ that reads in polar coordinates $\boldsymbol{r}=(r,\theta)$, 
\begin{equation}
\mathcal{H}=\frac{\hbar v_F}{\sqrt{2}}
\begin{pmatrix}
   0   &    \hat{L}_{-}    & 0  \\
  \hat{L}_{+}   &    0    & \hat{L}_{-}  \\
   0   &    \hat{L}_{+}    & 0     
  \end{pmatrix} + {V}(r),
\end{equation}
with the compact operator
$$\hat{L} = -ie^{i\tau\theta}\left(\partial_r + i\frac{\partial_\theta}{r}\right),$$
and ${V}(r)=V_0\Theta(R-r)$ the circular symmetric scalar-type scattering potential. It is evident that $[\mathcal{H}, \hat{\mathcal{J}}_z] = 0$ with the definition of $\hat{\mathcal{J}}_z = -i\hbar\partial_\theta + \hbar{S}_z$.  
As such, $\mathcal{H}$ acting on the spinor eigenfunctions of $\hat{\mathcal{J}}_z$ yields
\begin{equation}
\mathcal{H}\varphi_{l,s} = E\varphi_{l,s},
\end{equation}
where the wavefunctions $\varphi_l$ simultaneously satisfy $\hat{\mathcal{J}}_z\varphi_l = \hbar l\varphi_l$ with $l$ being an integer. After some standard derivations, we obtain for the conical bands (i.e. $s=\pm$)
\begin{equation}
\varphi_{l,s}^{(0,1)}(\boldsymbol{r}) =  \frac{1}{2\sqrt{\pi}}
\begin{pmatrix}
           h_{l-1}^{(0,1)}(qr)e^{-i\theta} \\
          i\sqrt{2}s h_{l}^{(0,1)}(qr) \\
          - h_{l+1}^{(0,1)}(qr)e^{i\theta}
\end{pmatrix}
e^{il\theta},   
\end{equation}
where $q=|E-{V}|/\hbar v_F$ and $s=\textrm{Sign}(E-{V})$. The radial function $h_l^{(0)}=J_l$ is the Bessel's function, and $h_l^{(1)}=H_l^{(1)}$ the Hankel's function of the first kind. The partial waves outside the scatterer ($r>R$) are therefore given by
\begin{subequations}
\begin{equation}
 \psi_{l,s}^{>}(\boldsymbol{r}) = \sqrt{\pi}i^{l-1}\left[\varphi_{l,s}^{(0)} + A_l\varphi_{l,s}^{(1)} \right],
\end{equation}
while inside the scatterer ($r<R$) the partial waves read
\begin{equation}
\psi_{l,s}^{<}(\boldsymbol{r}) = \sqrt{\pi}i^{l-1}B_l\varphi_{l,s'}^{(0)},
\end{equation}
\end{subequations}
where $A_l$ and $B_l$ denote the elastic partial wave reflection and transmission coefficients in the $l$ angular channel respectively. In order to obtain the explicit expressions of the partial wave coefficients, relevant boundary conditions (BCs) are needed.

\paragraph*{Boundary conditions.}
Recalling the commutation relations $[\hat{\mathcal{J}}_z, \mathcal{H}]=0$, we generally define a spinor wavefunction in polar coordinate
\begin{equation}\label{eq:wave} 
\psi(r,\theta) = [\psi_1, \psi_2, \psi_3]^{T} = \left(\begin{array}{c}
\mathcal{R}_1(r)e^{-i\theta} \\
\mathcal{R}_2(r) \\
\mathcal{R}_3(r)e^{i\theta}
\end{array}\right)e^{il\theta}, 
\end{equation}
that satisfies
\begin{equation} \label{eq:weq}
\mathcal{H}\psi=E\psi.
\end{equation}
Substituting Eq.~(\ref{eq:wave}) into the wave Eq.~(\ref{eq:weq}) and eliminating the angular components finally yield the following (one-dimensional first-order ordinary differential) radial equation
\begin{equation}\label{eq:radi}
\begin{aligned}
\frac{-i}{\sqrt{2}}\left(\begin{array}{ccc}
0 & \frac{d}{dr} + \frac{l}{r} & 0 \\
\frac{d}{dr} - \frac{l-1}{r} & 0 & \frac{d}{dr} + \frac{l+1}{r} \\
0 & \frac{d}{dr} - \frac{l}{r} & 0
\end{array}\right)\left(\begin{array}{c}
\mathcal{R}_1(r) \\
\mathcal{R}_2(r) \\
\mathcal{R}_3(r)
\end{array}\right) \\
=\frac{E - V(r)}{\hbar{v_F}}\left(\begin{array}{c}
\mathcal{R}_1(r) \\
\mathcal{R}_2(r) \\
\mathcal{R}_3(r)
\end{array}\right).
\end{aligned}
\end{equation}
Directly integrating the radial equation above over a small interval $r\in[R-\eta, R+\eta]$ defined around an interface at $r=R$ and then taking the limit $\eta\rightarrow 0$, we obtain
\begin{equation}\label{eq:rBCs}
\begin{aligned}
\mathcal{R}_2(R - \eta) &= \mathcal{R}_2(R + \eta), \\
\mathcal{R}_1(R - \eta) + \mathcal{R}_3(R - \eta) &= \mathcal{R}_1(R + \eta) + \mathcal{R}_3(R + \eta),
\end{aligned}
\end{equation}
provided that the potential $V(r)$ and the radial function components $\mathcal{R}_{1,2,3}(r)$ are all finite.
Reformulating such continuity conditions in terms of the corresponding wavefunction yields the BCs that we seek
\begin{equation}\label{eq:BCs}
\begin{aligned}
\psi_2^{<}(R,\theta) &= \psi_2^{>}(R,\theta), \\
\psi_1^{<}(R,\theta)e^{i\theta} + \psi_3^{<}(R, \theta)e^{-i\theta} &= \psi_1^{>}(R,\theta)e^{i\theta} + \psi_3^{>}(R,\theta)e^{-i\theta}.
\end{aligned}
\end{equation}

\paragraph*{Far-field solutions: $r\gg R$.}
Using the asymptotic form of the Hankel function $H_l^{(1)}(kr)\sim\sqrt{2/\pi kr}e^{i(kr-l\pi/2 - \pi/4)}$ and evaluating the outside wavefunction given in Eq.~(\ref{eq:OW}) at $r\gg R$, we arrive at
\begin{equation}\label{eq:fF2}
|\Psi_{s}^{\gg}(\boldsymbol{r})\rangle = e^{ikx}|\boldsymbol{k}_0,s\rangle + \frac{-i\sqrt{2/\pi k}\sum_lA_le^{il\theta}}{\sqrt{-ir}}e^{ikr}|\boldsymbol{k}_{\theta},s\rangle.
\end{equation}
It is evident from the Eq.~(\ref{eq:fF2}) and Eq.~(\ref{eq:fF1}) that
\begin{equation}\label{eq:fA}
f(\theta) = -i\sqrt{\frac{2}{\pi k}}\sum_{l}A_le^{il\theta}.
\end{equation}
Imposing relevant BCs given in Eq.~(\ref{eq:BCs}) on the total wavefunctions of both sides at the interface $r=R$, we have
\begin{equation}\label{eq:ABE}
\left\{
\begin{aligned}
B_lJ_l(qR) & = ss'\left[J_l(kR) + A_lH_l^{(1)}(kR)\right], \\
B_lX_{l}^{(0)}(qR) & = X_{l}^{(0)}(kR) + A_lX_{l}^{(1)}(kR), 
\end{aligned}\right.
\end{equation}
where $X_{l}^{(0,1)}=h_{l-1}^{(0,1)} - h_{l+1}^{(0,1)}$.
Solving the equation above, we finally determine
the unknown coefficients
\begin{equation} \label{eq:A}
A_l = -\frac{J_l(qR)X_{l}^{(0)}(kR)-ss' X_{l}^{(0)}(qR)J_l(kR)}{J_l(qR)X_{l}^{(1)}(kR)-ss' X_{l}^{(0)}(qR)H_l^{(1)}(kR)},
\end{equation}
and
\begin{equation} \label{eq:B}
B_l = \frac{H_l^{(1)}(kR)X_{l}^{(0)}(kR) - X_{l}^{(1)}(kR)J_l(kR)}{H_l^{(1)}(kR)X_{l}^{(0)}(qR) - ss'X_{l}^{(1)}(kR)J_l(qR)}.
\end{equation}

Using the basic relations of $J_{-l}=(-)^lJ_l$ and $H_{-l}^{(1)}=(-)^lH_l^{(1)}$, one can show the following symmetries
\begin{equation}
A_{-l}=A_{l}, B_{-l}=B_{l}.
\end{equation}
As such, the resulting probability density $\rho=\langle\Psi_{s}(\boldsymbol{r})|\Psi_{s}(\boldsymbol{r})\rangle$ and local current density $\boldsymbol{j}=\langle\Psi_{s}(\boldsymbol{r})|\hat{\boldsymbol{J}}|\Psi_{s}(\boldsymbol{r})\rangle$ can be calculated accordingly. In addition, the relevant scattering amplitudes $f(\theta)$ can be exactly obtained according to the Eq.~(\ref{eq:fA}) and hence related cross section given in Eqs.~(\ref{eq:TCS}) and (\ref{eq:TrCS}).

\begin{figure}[h]
\centering
\includegraphics[width=\linewidth]{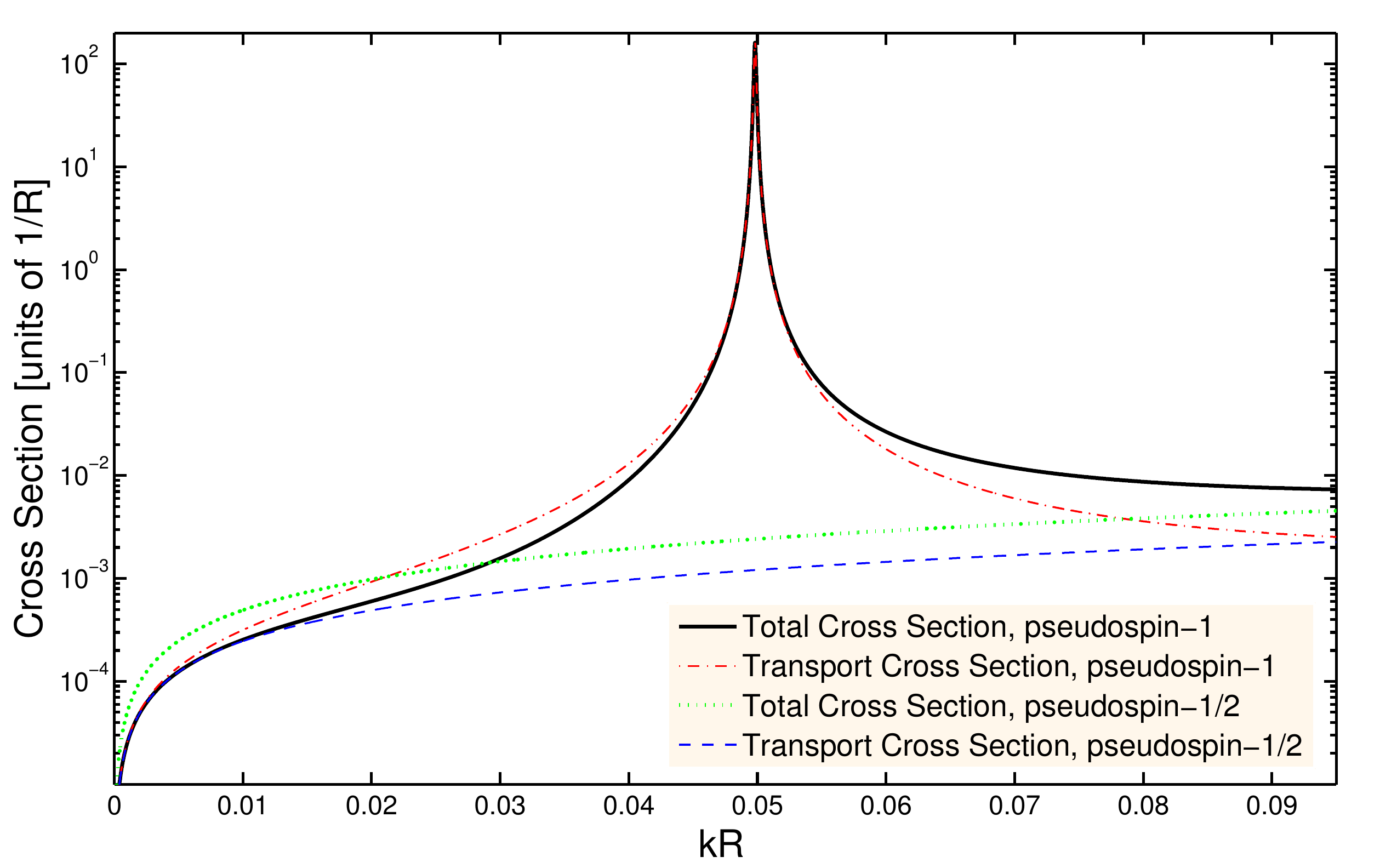}
\caption{Total and transport cross sections versus $x\equiv kR$ for a given weak scattering potential $\rho\equiv V_0R = 0.1$ in cases of pseudospin-$1$ and pseudospin-$1/2$ respectively.}
\label{fig:CSs}
\end{figure}

\paragraph*{Derivation of the Eq.~(\ref{eq:TrR}).}
By definition, the transport cross section can be obtained as 
\begin{equation}
\frac{\Sigma_{tr}}{R} = \frac{4}{x}\sum_{l=-\infty}^{\infty}\left\{|A_l|^2 
- \Re{[A_l(A_{l+1})^*]}\right\},
\end{equation} 
with $A_l$ being the reflection coefficients given in Eq.~(\ref{eq:A}). 
For $x\ll 1$, scattering is dominated by the lowest angular momentum 
channels $l=0,\pm1$. As a result, the transport cross section can be 
approximated as 
\begin{equation}\label{eq:aTr}
\frac{\Sigma_{tr}}{R}\approx \frac{4}{x}\left\{|A_0|^2 + 2|A_1|^2 
- 2\Re{[A_0(A_1)^*]}\right\},
\end{equation}
where
\begin{subequations}
\begin{equation}\label{eq:A0}
\begin{aligned}
A_0  &\approx -\frac{\pi x}{\pi x + i2\left[x\ln(\gamma_Ex/2) 
- J_0(\rho)/J_1(\rho)\right]} \\
&= -\frac{P_0}{P_0 + iQ_0},
\end{aligned}
\end{equation}
and
\begin{equation}\label{eq:A1}
\begin{aligned}
A_{\pm1} &\approx -\frac{\pi x^3}{\pi x^3 + i2\left[J_1(\rho)/J'_1(\rho) 
- x\right]} \\
&= - \frac{P_2}{P_2 + iQ_2},
\end{aligned}
\end{equation}
provided that the scattering potential is large, $\rho > 1$. 
\end{subequations}
Substituting the Eqs.~(\ref{eq:A0}) and (\ref{eq:A1})  into 
Eq.~(\ref{eq:aTr}), we obtain
\begin{equation}\label{eq:aaTr}
\begin{aligned}
\frac{\Sigma_{tr}}{R}\approx \frac{4}{x}&\left\{\frac{P_0^2}{P_0^2 + Q_0^2} 
+ 2\frac{P_2^2}{P_2^2 + Q_2^2} - \right. \\
& \left. 2\frac{P_0P_2Q_0Q_2}{(P_0^2 + Q_0^2)(P_2^2 + Q_2^2)}\right\}.
\end{aligned}
\end{equation}
Since $P_2 = x^2P_0=\pi x^3\ll1$, the transport cross section will approach  
$\Sigma_{tr}/R\sim 4/x~(8/x)$ about $Q_0=0~(Q_2=0)$, and 
$\Sigma_{tr}/R\sim x\ll1$ otherwise. It is thus reasonable to 
reduce Eq.~(\ref{eq:aaTr}) to
\begin{equation}\label{eq:aaaTr}
\frac{\Sigma_{tr}}{R} \approx \frac{4}{x}\left\{\frac{P_0^2}{P_0^2 + Q_0^2} 
+ 2\frac{P_2^2}{P_2^2 + Q_2^2}\right\},
\end{equation}
where $Q_0 = 2[x\ln(\gamma_Ex/2) - J_0(\rho)/J_1(\rho)]$ and 
$Q_2=2[J_1(\rho)/J'_1(\rho) -x]$. Since they are from different terms
and each term has considerable values near $Q_0=0$ or $Q_2=0$, we can
expand $Q_0$ and $Q_2$ about the zeros of $J_0(\rho)$ and $J_1(\rho)$,
respectively, to get
\begin{subequations}
\begin{equation}
Q_0 \approx 2\left[\rho - \rho_{0,m} + x\ln\frac{\gamma_Ex}{2}\right],
\end{equation}
and
\begin{equation}
Q_2 \approx 2(\rho - \rho_{1,n} - x),
\end{equation}
with $m, n=1,2,3,\cdots$ and $\rho_{0,m}$ and $\rho_{1,n}$ denoting the 
$m$th and $n$th zeros of the Bessel functions $J_0$ and $J_1$, respectively.
\end{subequations}
Substituting these into the Eq.~(\ref{eq:aaaTr}), we arrive at Eq.~($4$) 
in the main text.

\section{Characterizing superscattering with total cross section}

The total cross section $\Sigma$ is an alternative quantity to 
characterize superscattering, with consistent results as the transport 
cross section, as shown in Fig.~\ref{fig:CSs}. In particular, from the 
total cross-section curves (black and green), one can infer the same 
scattering behaviors as from the transport cross section. In fact, with
the definition of total cross section: $\Sigma/R = 4/x\sum_{l}|A_l|^2$, 
we can obtain a closed-form formula in the weak scattering potential regime:
\begin{equation}
\frac{\Sigma}{R} \approx \frac{\pi^2}{4}\rho^2x\left[1 + \frac{8x^2}{\pi^2(\rho-x)^2x^4 + 16(\rho-2x)^2}\right].
\end{equation}
For $\rho=2x$, the total cross section gives the same resonant peak value 
$\sim16/\rho$ as the transport cross section would.


%
\end{document}